\documentstyle[times,psfig]{jaa}
\begin{document}
\title[Theoretical studies]{Asteroseismic theory of rapidly oscillating Ap stars} 
\author[M.S. Cunha]%
       {Margarida S. Cunha\thanks{e-mail:mcunha@astro.up.pt} \\ 
        Centro de Astrof\'\i sica da Universidade do Porto,Rua das estrelas,4150,Porto, Portugal}
\maketitle
\label{firstpage}
\begin{abstract}
The present paper reviews some of the important advances made over the last decade concerning theory of roAp stars. 
\end{abstract}

\begin{keywords}
stars: variable, Chemically peculiar, roAp, magnetic
\end{keywords}
\section{Introduction}
\label{sec:intro}
The success of the asteroseismic studies of rapidly oscillating Ap (roAp) stars depends strongly on our ability to understand their oscillation spectra. Questions like {\em which modes are excited and why}, {\em what is the expected spacing between eigenfrequencies}, {\em how many components are expected to be found in the multiplet structures}, or {\em what are the relative amplitudes of the different components of the multiplets} have been central to recent theoretical studies of roAp stars. In this paper we will review different aspects of the theoretical work recently carried out in this field.

\section{From spherical symmetry to roAp stars}

A key aspect to have in mind when studying pulsations in roAp stars is the presence of different physical agents which influence the pulsations in a non-spherically symmetric way.  The deviations from spherical symmetry add additional complexity to the interpretation of the oscillation spectra of these stars, as well as to their asteroseismic study in general. 
Through the rest of this section we will attempt to summarize some of the properties of the eigenfrequencies and eigenfunctions of modes of oscillation with fixed radial order $n$ and fixed degree $l$, in models with different symmetry properties, without going through the cumbersome calculations needed to derive them. More information on the subject can be found in e.g.\ Unno et al.\ (1989) and Gough (1993), and, for the specific case of roAp stars, in Bigot and Dziembowski (2002) and Gough (2003). We will consider linear, adiabatic pulsations. Moreover, we will assume that the impact of the symmetry-breaking agents on pulsations is sufficiently small that their effects can be superposed linearly. Some of the non-axisymmetric agents present in roAp stars, particularly the magnetic field, can distort the modes of oscillation to an extent that a single eigenmode will no longer be associated to a single value of $l$ (Dziembowski and Goode 1996). However, the implications of this fact will be postponed to the next section. Thus, for the rest of this section it will be assumed that each eigenmode is well represented by a single degree value $l$. 

\begin{figure}
\centerline{\psfig{file=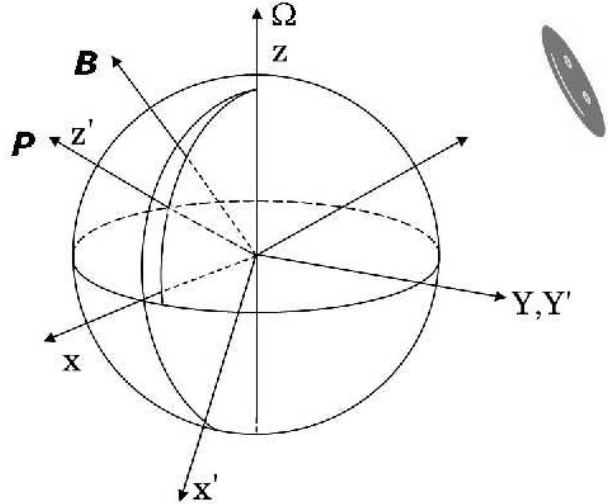,width=8.5cm}}
    \caption[]{Schematic view of the coordinate systems that are most relevant to the study of pulsations in roAp stars (see text). }		
\label{fig:frame}	
\end{figure}

The modes of oscillation in the models that will be considered are described by the eigenvalue equation $\omega^2\psi=\mathcal{A}(\psi)$, where $\omega$ are the eigenfrequencies, $\psi$ are scalar eigenfunctions (e.g. radial component of the displacement) and $\mathcal{A}$ is a generic differential operator which takes a different form depending on the model considered.
 
In a spherically symmetric model the choice of direction for the polar axis of our coordinate system is arbitrary: all equilibrium quantities in the differential operator $\mathcal{A}$ are independent of that choice. Since the orientation does not influence the eigenvalue problem, the $2l+1$ eigenfrequencies associated with a mode of radial order $n$ and degree $l$ will be degenerate.  The angular parts of the corresponding $2l+1$ eigenfunctions are described by arbitrary (independent) linear combinations of the $2l+1$ spherical harmonics $Y_l^m$ (with the azimuthal order $m$ varying from -$l$ to $l$). In particular, one can define a set of $2l+1$ independent eigenfunctions $\psi^0_m$ for the spherical symmetric model, such that each is proportional to a different spherical harmonic function $Y_l^m$, with $l$ fixed.

Next consider the case in which the spherical symmetry of the problem is broken due to an axisymmetric effect produced by a given physical agent, such as an axisymmetric magnetic field, or rotation. 
In this case $\mathcal{A}$ incorporates the spherically symmetric differential operator referred above plus additional terms associated with the direct and indirect effects of the non-spherically symmetric agent on the oscillations.  Because of the lack of spherical symmetry, the orientation of the eigenmodes is no longer arbitrary. If we use the axis of symmetry of the symmetry-breaking agent as the polar axis $(\theta=0)$, in a spherical coordinate system (r,$\theta$,$\phi$) attached to the star, the additional effects incorporated in $\mathcal{A}$ will be independent of $\phi$. We can then have two situations: if the non-axisymmetric effect does not depend on the sense of the axis of symmetry (e.g.\ the Lorentz forces generated by the perturbed magnetic field), then the problem is invariant under reflection about the equator and some degeneracy remains. If, on the other hand, the effect depends on the sense of the axis of symmetry (e.g.\ the Coriolis force produced by the rotation of the star), then the degeneracy is completely lifted.
In the first case, all but one of the $2l+1$ eigenfrequencies will be degenerated in pairs, and the angular part of the corresponding pairs of eigenfunctions will be arbitrary (independent) linear combinations of the pairs of functions $Y_l^{-m}$ and $Y_l^m$. Moreover, the angular part of the eigenfunction associated with the 'unpaired' eigenfrequency will be given by $Y_l^0$.
In the second case the $2l+1$ eigenfrequencies will all be different. Moreover, the arbitrariness in the angular part of the eigenfunctions will be totally removed: the $2l+1$ eigenfunctions will each have an angular part described by a spherical harmonic function $Y_l^m$ of a different azimuthal order $m$.  
Thus, if the perturbation is axisymmetric, we can still define a set of $2l+1$ independent eigenfunctions $\psi^1_m$ that are solutions to the problem, each proportional to a different spherical harmonic function $Y_l^m$. 

In roAp stars there are several symmetry-breaking agents which simultaneously influence the oscillations. This is the case of rotation, magnetic fields and possible structural differences associated with the spots observed at the surface of these stars. The fact that the axes of symmetry of these agents (assuming they are all axisymmetric), are not generally aligned, increases the complexity of the problem. To proceed with this description, let us specify the symmetry-breaking agents in our problem, by adopting the traditional oblique rotator model: a model of a magnetic rotating star, in which the axis of symmetry of the axisymmetric magnetic field is inclined in relation to the rotation axis.
 
If the magnetic field is force free (as is commonly assumed in the theoretical works of roAp stars), it will essentially affect the oscillations directly, via the Lorentz force, associated with the perturbed magnetic field. Moreover, according to Bigot and Dziembowski (2002) the most important effects of rotation regarding the pulsations of roAp stars are the effect of the Coriolis force and the effect of the centrifugal distortion of the star. In fact they point out that, due to the high radial order of the oscillations observed in roAp stars, the latter, despite being of second order in the angular velocity $\Omega$, is generally larger than the former. An essential difference between the effect of the Coriolis force and the effects of the Lorentz force and centrifugal distortion is that only the first of these can detect the sense of the corresponding axis of symmetry. As shown in Bigot and Dziembowski (2002), this property of the Coriolis force has important observational consequences, even when the effect of the latter is small compared with the other two effects. 

Since the axes of symmetry of the effects considered are not aligned, if we chose our coordinate system such that the polar axis is aligned with one of these axes, the effect produced by the other symmetry-breaking agent will depend on both $\theta$ and $\phi$. Hence, the angular parts of the $2l+1$ eigenfunctions will be given by different (but not arbitrary) linear combinations of the $2l+1$ functions $Y_l^m$ defined in that coordinate system. Generally there will be no degeneracy in the solutions. The $2l+1$ eigenfrequencies will be all different and the linear combinations that define the angular parts of the eigenfunctions will be all well determined. Hence, unlike before, it is generally not possible to associate each eigenfunction with an individual value of $m$ in either of the two coordinate systems. 

To simplify our discussion further, let us consider the specific case of dipole modes (i.e.\ $l=1$). If we were to neglect the effect of the Coriolis force, then the problem would be invariant under the transformation $m\rightarrow -m$. Thus, the coefficients of the terms $Y_1^1$ and $Y_1^{-1}$ in the linear combinations that define the angular part of the eigenfunctions would necessarily have the same absolute value. In fact, if there were no Coriolis force, a coordinate system would exist (with polar axis indicated by ${\bf\it P}$, in figure \ref{fig:frame}) such that one of the eigenfunctions would be axisymmetric about the polar axis (i.e.\ its angular part would be proportional to $Y_1^0$) while the other two eigenfunctions would each be axisymmetric about one of the other two axis of the coordinate system (i.e.\ their angular parts in that coordinate system would be proportional, respectively, to $Y_1^{-1}+Y_1^{1}$ and $i(Y_1^{-1}-Y_1^{1})$). The polar axis of that coordinate system is in the plane defined by the magnetic and rotation axes, and its position in that plane depends on the relative strengths of the magnetic and centrifugal effects. 

The effect of the Coriolis force on the pulsations of roAp stars is typically small compared with the magnetic and the centrifugal effects. However, unlike the other effects, the former generates an asymmetry between the absolute values of the $m=1$ and $m=-1$ coefficients of the linear combinations that determine the eigenfunctions (Bigot and Dziembowski 2002). As a consequence, the pulsation axes of the three eigenfunctions, which in the absence of the Coriolis force would be directed along the axes of the reference frame described above, change direction, over the period of pulsation, in planes which are perpendicular to the plane defined by the rotation and magnetic axes. The maximum of the radial displacement vector of each eigenmode describes an ellipse, over the corresponding period of pulsation, in the corresponding plane.

When the Coriolis effect is not neglected the eigenmodes can alternatively be describe in a precessing reference frame (Gough 2003). The polar axis of that reference frame is slightly deviated from the plane defined by the rotation and magnetic axes and precesses about the polar axis of the pulsation coordinate system defined in the absence of the Coriolis effect (${\bf\it P}$ in figure \ref{fig:frame}), on a long time scale. The eigenmodes of the problem defined in this precessing frame correspond to specific combinations of the eigenmodes that are solution to the problem in the reference frame attached to the star. In the precessing reference frame one of the eigenmodes will be axisymmetric about the polar axis while the other two linearly independent modes will precess in a plane perpendicular to it.

\section{The observer's view}

Viewed from the earth the eigenfunctions will be proportional to linear combinations of the type $\sum_{m=-l}^{l}A_{l,m}\cos(\omega -m\Omega)t$, where $\Omega$ is the angular velocity of the star and $t$ is the time.
Thus, each of the $2l+1$ eigenmodes will generally be seen as a $(2l+1)$-component multiplet. Because of the effect of the Coriolis force, the coefficients $A_m$ and $A_{-m}$ will generally have different absolute values. This inequality of the side peaks associated with the same value of $|m|$ depends on the relative importance of the Coriolis force and on the orientation of the mode plane. Moreover, the ratio between the sum of the pairs of side peaks to the central peak depends both on the inclination of the rotation axis to the line of sight and on the orientation of the mode plane.
\begin{figure}
\centerline{\psfig{file=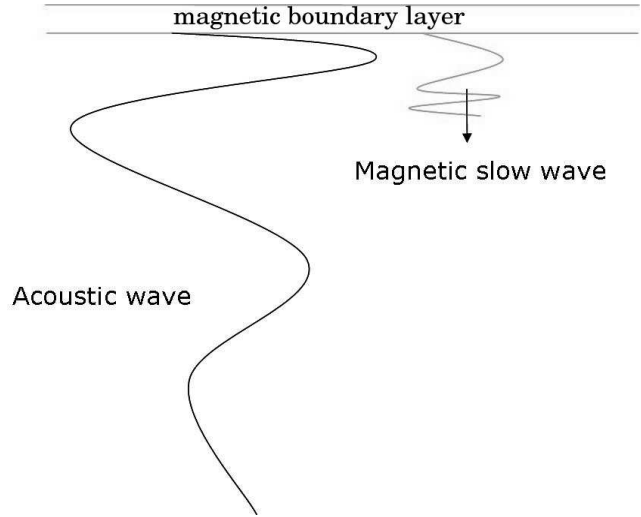,width=8.5cm}}
    \caption[]{Schematic representation of the magnetic effect on the oscillations of roAp stars. There is a region in the magnetic boundary layer where the waves are magnetoacoustic. In the interior, the magnetic and acoustic parts decouple and the magnetic component is expected to dissipate due to the rapid increase in its wavenumber. }
\label{fig:waves}
\end{figure}

\begin{figure}
\centerline{\psfig{file=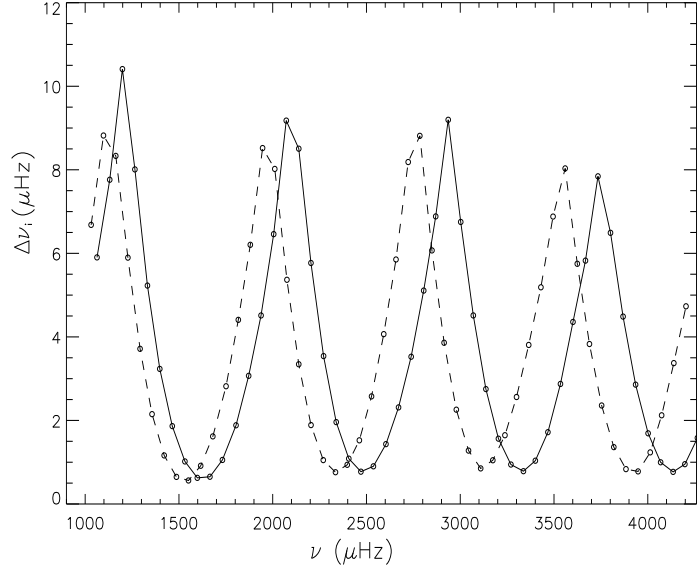,width=10.0cm,height=8.0cm}}
    \caption[]{Imaginary part of the magnetic perturbations to the oscillations in two polytropic models of a roAp star: solid line - mass $M=2.0{\rm M}_{sun}$ and radius $R=1.9{\rm R}_{sun}$; dashed line - mass $M=1.5{\rm M}_{sun}$ and radius $R=1.75{\rm R}_{sun}$.}			
\label{fig:dwi}
\end{figure}

As mentioned in the previous section, the magnetic field can distort the modes of oscillation to an extent that a single mode might no longer be well represented by a single spherical harmonic of degree $l$. Even though the overall effect of the magnetic field on the oscillations is small, there is as region, near the surface of the star, where the magnetic effect cannot be regarded as a small perturbation. It is this region, where the magnetic pressure is comparable or larger than the gas pressure, that is responsible for the significant distortion of the eigenmodes. In the previous section this distortion of the eigenfunctions from single spherical harmonics was neglected. However, the latter has been studied by several authors (e.g.\ Dziembowski and Goode 1996, Saio and Gautschy 2004) in a context in which only the magnetic effect was accounted for (i.e.\ the effect of rotation on the oscillations was neglected). In this case the angular part of the eigenfunctions are represented, in a coordinate system with the polar axis along the magnetic axis, by sums of spherical harmonics of different degree $l$ and the same azimuthal order $m$, i.e.\ $\sum_{l} B_lY_l^m$. In the observer's reference frame, each of the individual $l$ components of the linear combination that describes the angular part of a mode of frequency $\omega$ will give rise to a $(2l+1)$-component multiplet of the type $\sum_{m=-l}^{l}b_{l,m}\cos(\omega -m\Omega)t$. Thus, as seen by the observer, an eigenmode will be associated with a $(2l_{max}+1)$-component multiplet, where $l_{max}$ is the maximum $l$ value such that the corresponding coefficient $B_{l_{max}}$, taken at the surface of the star, is still sufficiently large for that component to be observable after being averaged over the visible disk. Note that because the rotational effect on the oscillations has been neglected in these studies, the absolute value of the coefficients $b_{l,m}$ and $b_{l,-m}$ obtained in the latter is the same.

\section{Basic signatures} 

In order to interpret the power spectra of roAp stars we need also to understand how the eigenfrequencies are modified by the non-axisymmetric effects described in the previous sections. In particular, it is most relevant to learn how basic signatures used in asteroseismic studies, like the small and large separations are affected. In fact, the large separations of roAp stars have been compared to theoretical predictions by Matthews et al.\ (1999) and Cunha et al.\ (2002), using standard spherically symmetric models, and in both cases discrepancies between theory and observations were found.  Cunha et al.\ (2002) have attempted to modify some of the parameters included in the models used for a particular roAp star (like chemical abundance and amount of overshooting in the core), but the best agreement the authors found implied values for the metal abundance that were too low to be acceptable. Even though they could reconcile the theoretical results with the observations when the error-bars in the luminosity and effective temperature of the star were taken into account, the fact that the discrepancy between the theoretical predictions and the observations is systematic for roAp stars (Matthews et al.\ 1999) might indicate that some important physics was missing in the spherically symmetric models used.

 The effect of the magnetic field on the frequency of the oscillations in roAp stars  has been studied by different authors who accounted for the fact, first noted by Biront et al.\ 1982), that in the surface layers, where the gas pressure is low, the magnetic field effect cannot be treated as a small perturbation (e.g. Dziembowski and Goode 1996, Bigot et al. 2000, Cunha and Gough 2000,Cunha 2001, Saio and Gautschy 2004).  

In the region where the gas and magnetic pressure are comparable, the waves are magnetoacoustic. Where the gas pressure dominates, however, the latter decouple into magnetic and acoustic components. Due to the rapid increase of the magnetic wavenumber, as the density gets larger, it is expected that the magnetic waves will dissipate (Roberts and Soward 1983). Therefore, the acoustic wave in the interior is continuously losing energy through the coupling that takes place in the magnetic boundary layer (see figure \ref{fig:waves}). As a consequence of this energy loss, the eigenfrequencies are complex, even when the oscillations are studied in an adiabatic approximation and are fully reflected at the surface.   Figure \ref{fig:dwi} shows the imaginary part of the frequency perturbations, which originate from the coupling described above, in a polytropic model of a roAp star. It is clear from the figure that there are frequency intervals for which the dissipation is very large, when compared with other frequency intervals for which the latter is negligible. This fact might influence the frequency range in which oscillations are excited in a given roAp star. Thus, magnetic coupling might introduce mode selection in some roAp stars.

\begin{figure}
\centerline{\psfig{file=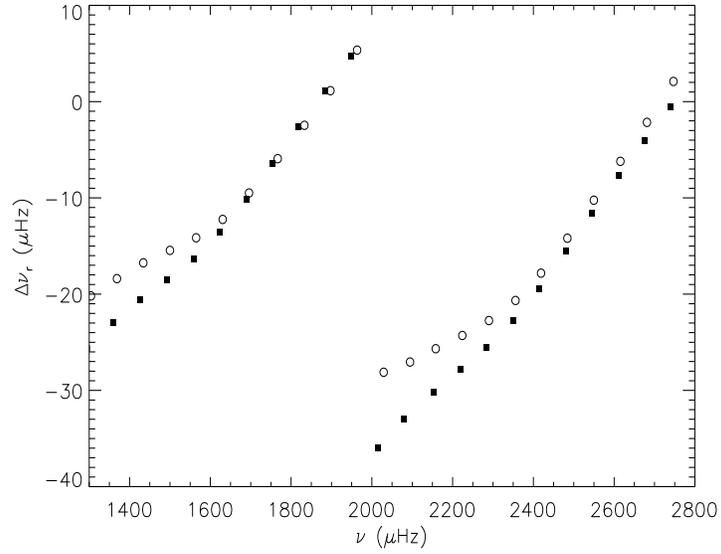,width=10.0cm,height=8.0cm}}
    \caption[]{Real part of the magnetic perturbations to the oscillations of a polytropic model with a mass $M=1.5{\rm M}_{sun}$ and a radius $R=1.75{\rm R}_{sun}$. The open circles and filled squares show the perturbations for modes of degree $l=1$ and $l=3$, respectively.}			
\label{fig:dw}
\end{figure}

\begin{figure}
\centerline{\psfig{file=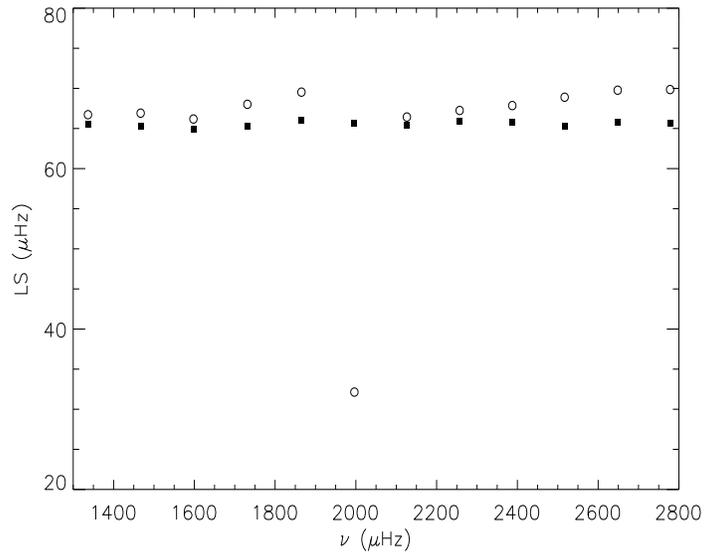,width=10.0cm,height=8.0cm}}
    \caption[]{Magnetic perturbations to the large separations of the model described in the caption of figure \ref{fig:dw}. The perturbed large separations (shown by open circles) are compared with the unperturbed ones (shown by filled squares).}			
\label{fig:ls}
\end{figure}

Due to the high radial order of the oscillations observed in roAp stars, one might expect, to first approximation, to find the eigenfrequencies regularly spaced in the oscillation spectra. The large separations are defined as the difference in frequency between modes of the same degree $l$ and consecutive radial orders, $n$. Thus, if modes of both even and odd degree are excited, the large separations will correspond to approximately twice the average spacing between consecutive modes in the oscillation spectra. Moreover, modes of degree $l$, and $l+2$ and, respectively, radial orders $n$ and $n-1$ are almost degenerated in frequency. The difference between their frequencies is known as the small separations and in roAp stars they are typically only a few $\mu$Hz.

In figure \ref{fig:dw} the real part of the magnetic perturbations to the eigenfrequencies of a polytropic model of a roAp star are shown for modes of degree $l=1$ and $l=3$. The dependence of the effect on the degree of the mode is clearly seen in some frequency domains. Thus the small separations will be significantly modified from the value they would have in the absence of a magnetic field.
In figure \ref{fig:ls} the perturbations to the eigenfrequencies of the same model are combined, for the mode of degree $l=1$, to show the magnetic perturbations to the large separations. As can be seen, the large separations are generally slightly increased by the effect of the magnetic field. However, it is also apparent from the figure that at some particular frequencies the large separations can be significantly modified. Thus, when inspecting the oscillation spectra of some roAp stars one might expect to find anomalies in the usual mode spacing, resulting from these abrupt changes in large separations at given frequencies.

\section{Conclusion}

The theory of rapidly oscillating Ap stars has progressed considerably over the past decade. Some of the major advances concern the effect of the magnetic field on the oscillations, and the combined magnetic and rotational effects on the orientation of the modes. As argued in this review, both of these issues have important implications to the interpretation of the oscillation spectra of roAp stars. However, further progress is desirable, particular in what concerns a consistent theory that takes into account the combined effect of the magnetic field and rotation, without neglecting the distortion of the eigenmodes from single spherical harmonics. Additionally, progress is needed in what concerns the modelling of the mechanism by which the oscillations are reflected near the surface of roAp stars. Approximations like introducing a mechanical boundary condition that assures full reflection of the oscillations near the surface, or assuming a locally uniform magnetic field, can disguise the true reflection process. Since the large separations depend on the size of the pulsation cavity, further improvements on the modeling of the layers where reflection takes place would also be important.

This paper was centered on adiabatic studies of pulsations in roAp stars. However, progress has also been made concerning important theoretical issues that relate to non-adiabatic studies. In particular, it is believed today that oscillations in roAp stars are excited by the $\kappa$-mechanism in the hydrogen ionization region (Dziembowski and Goode 1996). High frequency oscillations are found unstable in models of roAp stars if the surface convection is assumed to be suppressed by the magnetic field at least in some angular region of the star (Balmforth et al.\ 2001, Cunha 2002), or if it is assumed that the star has a chromosphere (Gauthschy et al.\ 1998). A theoretical instability strip is now available (Cunha 2002), which can be used to compare with the observations, as well as to motivate further observations of Ap stars that populate regions of luminosity and effective temperature where up to date no high frequency oscillations were found.

\section{Acknowledgments}

MC is supported by FCT-Portugal, through the grants BPD/8338/2002 and 
POCTI/CTE-AST/57610/2004.

\end{document}